# Equilibrium Properties of Mixtures of Bosons and Fermions

U. F. Edgal[*] and D. L. Huber[‡]

*Department of Engineering Technology, Old Dominion University*
*Norfolk, Virginia, 23529*

*and*

*Department of Physics, University of Wisconsin-Madison*
*Madison, Wisconsin, 53706*

[*] To whom correspondence should be addressed. Department of Engineering Technology, Old Dominion University, 5115 Hampton Blvd., Norfolk, VA 23529. Phone: (757)683-6718. Fax: (757)683-5655. Email: uedgal@odu.edu.
[‡] University of Wisconsin-Madison

# ABSTRACT

Partial Quantum Nearest Neighbor Probability Density Functions (PQNNPDF's) are formulated for the purpose of determining the behavior of quantum mixed systems in equilibrium in a manner analogous to that provided for classical multi-component systems. Developments in partial quantum m-tuplet distribution functions, a generalization of the partial quantum radial distribution function, along with their relationship to PQNNPDF's, are briefly elucidated. The calculation of statistical thermodynamic properties of quantum mixtures is presented for arbitrary material systems. Application to the limiting case of dilute, weakly correlated quantum gas mixtures has been outlined and the second virial coefficient is derived. The case of dilute strongly degenerate mixtures is also addressed, providing an expression for the PQNNPDF applicable in this thermodynamic regime.

# I. Introduction

The present paper addresses the most general type of material system, quantum mixed systems. It extends developments recently provided for the equilibrium microstructure and statistical thermodynamic properties of multi-component systems of the classical type[1] to those of the quantum type. This is similar to the kind of extension that was done for pure classical systems[2] to pure quantum materials[3]. The present investigation therefore begins by re-expressing the "sum of states" for mixed quantum systems (with an arbitrary number of Fermion and Boson species) as a multidimensional integral over a portion of all of classical phase space. Once again, the concepts of generalized order, the reduced one particle phase space, and the partial nearest neighbor probability density function (PNNPDF) are found to be essential ingredients (or analytic tools) in the formulation. These tools are new and are the subject of vigorous investigation. The present scheme of this paper addresses quantum systems involving an arbitrary mix of Fermions and Bosons for which in the past, various methods have usually been restricted to special situations where corrections to or deviations from classical or other limiting results are evaluated approximately[4]. In this regard, semi-classical analytical expansion schemes (for pure and mixed virial coefficients, direct and exchange contributions to correlation functions, etc.) are by far the most predominant methods. Much of the currently available advanced theoretical methods[5] have not usually been applied to multi-component systems even though a variety of such systems[5,6] have been discussed in condensed matter physics. Recently, there has been revived interest in mixtures of Bose and Fermi gases comprised of trapped cold atoms[7]. In comparison with solids, such atomic systems (with their controllable densities and comparatively well known inter-atomic interactions), may provide the definitive tests of various theories of quantum gas mixtures.

Without loss of generality, we employ the coordinate representation and thus express the trace (Tr) of the Boltzmann operator for a multi-component system of $N_i$ particles of species i (i = 1, ..., n) in a volume V as:

$$\tilde{Z}(N_1,\cdots,N_n,V) = Tr\left\{\exp\left(-\beta H_{N_1,\ldots,N_n}\right)\right\} = \frac{1}{\prod_{i=1}^{n}\left(N_i!\sigma_i^{N_i}\right)} \int\cdots\int W_{N_1,\ldots,N_n} dX''_{k''_1 1} \cdots dX''_{k''_N N}$$

(1)

$\beta = 1/kT$, k is Boltzmann constant, T is the temperature, $H_{N_1,\ldots,N_n}$ is the Hamiltonian operator of a system having n different species of particles, with the i-th species involving $N_i$ particles. We have that $N_1 + \ldots + N_n = N$, and partial densities involve $\rho_i = \frac{N_i}{V}$ (with the overall particle density being $\rho = \frac{N}{V}$). The constants $\sigma_i$ make $\tilde{Z}$ dimensionless, and $X''_{k''_1 1},\cdots,X''_{k''_N N}$ specify a given microstate in classical phase space. The first subscripts $k''_i$ of the $X''_{k''_i i}$ variables refer to species type, and thus $k''_i$ may assume any of the values from 1 to n. (Hence $N_i$ of the $k''_i$ subscripts have the value i). For simplicity, internal degrees of freedom are not rigorously treated, providing a strictly non-relativistic treatment where spin and other internal degrees of freedom may be introduced in the usual semi-classical manner. In the simplest case where the portion of phase space over which the integral is specified is configuration space, the phase space distribution is the "Multi-component Slater Sum" written as:

$$W_{N_1,\ldots,N_n}(X''_{k''_1 1},\cdots,X''_{k''_N N}) = \prod_{i=1}^{n}(N_i!\lambda_i^{3N_i})\sum_{\alpha_1,\ldots,\alpha_n}\psi^*_{\alpha_1,\ldots,\alpha_n}(X''_{k''_1 1},\cdots,X''_{k''_N N})e^{-\beta H_{N_1,\ldots,N_n}}\psi_{\alpha_1,\ldots,\alpha_n}(X''_{k''_1 1},\cdots,X''_{k''_N N})$$

(2)

$\sigma_i = \lambda_i^3$. The wave function $\psi_{\alpha_1,\ldots,\alpha_n}$ is rewritten (with some rearrangement of the coordinates) as the product

$$\psi_{\alpha_1,\ldots,\alpha_n}(X''_{11},\cdots,X''_{1N_1};\cdots;X''_{n1},\cdots,X''_{nN_n}) = \prod_{i=1}^{n}\left(\psi_{\alpha_i}(X''_{i1},\cdots,X''_{iN_i})\right)$$

(3)

$\{\psi_{\alpha_i}\}$ is a complete set of orthonormal wave functions symmetrized or antisymmetrized according to whether the i-th species involves Bosons or Fermions. The asterisk * implies complex conjugation, $\lambda_i$ is the thermal wavelength $\left(h^2/2\pi m_i kT\right)^{1/2}$ of the i-th species, h is Planck's constant, and $m_i$ is mass of species i particle. $X''_{i1},\cdots,X''_{iN_i}$ refer to the translational coordinates of species i particles. $W_{N_1,\ldots,N_n}$ is the diagonal element of the Boltzmann operator in coordinate representation. This function is the thermal Greens function (also the complex time propagator) which can be rewritten as

$\left(\prod_{i=1}^{n}N_i!\lambda_i^{3N_i}\right)\left\langle X''_{k''_1 1},\cdots,X''_{k''_N N}\left|e^{-\beta H_{N_1,\ldots,N_n}}\right|X''_{k''_1 1},\cdots,X''_{k''_N N}\right\rangle$ showing its analogy with the

Greens function of quantum field theory.

Like the single component case[3], the product property of the Multi-component Slater Sum is also well acclaimed (where for instance, it provides some basis for the Ursell cluster expansion scheme for multi-component quantum systems)[4]. The Multi-component Slater Sum product property is stated as $W_{N_1,\ldots,N_n} \approx W_{m_1,\ldots,m_n}W_{N_1-m_1,\ldots,N_n-m_n}$, where particle coordinates are separated into two groups, and those used to compute $W_{m_1,\ldots,m_n}$ are distinct from those used to compute $W_{N_1-m_1,\ldots,N_n-m_n}$. Any two coordinates $X''_{k''_i i}$, $X''_{k''_j j}$ belonging to different groups must satisfy (similar to the single component case)[3] the separation condition $r_{k''_i k''_j}, \lambda_{k''_i}, \lambda_{k''_j} \ll \left|X''_{k''_i i} - X''_{k''_j j}\right|$. ($r_{k''_i k''_j}$ is the effective inter-particle interaction range between particles of species $k''_i$ and $k''_j$). Extensive discussions of the above product property is available in the literature,[3,4] but this has been much more so for the single component case than for the multi-component case. In the developments of this paper, the two groups of particle coordinates of interest involve the m nearest neighbors of the "origin" (said to be randomly situated within volume V) on the one hand (where $m_i$ particles are of species i for i = 1, ... , n, and thus $m_1 + ... + m_n = m$) and the rest of the $[(N_1 - m_1), ... , (N_n - m_n)]$ particles on the other hand. As before[3], since the "surface-to-volume ratio" tends to zero as volume increases, we can always find some m which is finite but large enough (probably a few thousands or so, typically), such that the separation condition stated above may hold for an overwhelming set of coordinate pairs. (The exception being the relatively small number of coordinate pairs involving particles close to the boundary separating the volume containing the m nearest neighbors and the rest of the system particles).

However, the much milder form of the product property is once again what is actually needed. Noting that we may rigorously write the exact equality $W_{N_1,...,N_n} = W_{m_1,...,m_n} W_{N_1-m_1,...,N_n-m_n} F_{m_1,...,m_n}$ (where $F_{m_1,...,m_n}$ is some number), and if we may also write $W_{m_1,...,m_n} = e^{w_{m_1,...,m_n}}$, $W_{N_1-m_1,...,N_n-m_n} = e^{w_{N_1-m_1,...,N_n-m_n}}$, $F_{m_1,...,m_n} = e^{f_{m_1,...,m_n}}$, we need $|w_{m_1,...,m_n}|, |w_{N_1-m_1,...,N_n-m_n}| \gg |f_{m_1,...,m_n}|$ for m, (N – m) >> 1. (The original product property requires $f_{m_1,...,m_n} \approx 0$). This is the first statement of the milder form of the product property. The second statement of the milder form of the product property requires that $f_{m_1,...,m_n} - f_{m_1-m'_1,...,m_n-m'_n} \to 0$ as $m_i$ gets larger ($m_i > m_i$' and $m_i$' is fixed; i = 1, ... , n). This limiting behavior is replaced by an equality for all allowed m values in the original statement of the product property. (Clearly, "often enough", each $m_i$ increases for any considerable increase in the number of nearest neighbors considered in real homogeneous systems). Hence for m sufficiently large (m > m'), the mild form of the product property allows us write:

$$\frac{W_{N_1,...,N_n}}{W_{N_1-m'_1,...,N_n-m'_n}} = \frac{W_{N_1-m_1,...,N_n-m_n} W_{m_1,...,m_n} F_{m_1,...,m_n}}{W_{N_1-m_1,...,N_n-m_n} W_{m_1-m'_1,...,m_n-m'_n} F_{m_1-m'_1,...,m_n-m'_n}}$$

$$\approx \frac{W_{m_1,...,m_n}}{W_{m_1-m'_1,...,m_n-m'_n}} = e^{w_{m_1,...,m_n} - w_{m_1-m'_1,...,m_n-m'_n}}$$

(4)

The logarithmic nature of the milder form of the product property implies much less stringent requirement on how large m needs to be. Hence while the arguments of the surface-to-volume ratio requires m to be at least a few thousands or so for Eq.(4) to be valid, the milder form of the product property requires m to be much smaller (probably a few tens or so). In the very close neighborhood of zero temperature where $\lambda_i \to \infty$ however, it is expected that m will need to be prohibitively large; implying our scheme may not be used right down to zero temperature. Also, some revision of the product property and thus of our overall approach may be required in systems with strong density fluctuations (as in phase transition regions), as well as systems in which phase correlations between particles extend over macroscopic distances ( as in superfluids and superconductors). It is again however interesting to observe that the product property allows us avert the phase (or sign) problem usually encountered in calculations involving Fermion systems.

In section II, the concepts of generalized order and the reduced one particle phase space are employed identically as in the classical case[1,2] and the single component quantum case[3] to develop the statistical thermodynamic formalism for quantum mixed systems. It is especially noted that the formalism applies not only for macroscopic systems, but also for all system scales. Also introduced in section II, is the notion of the partial Quantum NNPDF (or PQNNPDF) which is the analog of PNNPDF's in classical multi-component systems[1]. In section III, further exposition of microstructure of quantum systems is briefly provided, developing PQNNPDF's, and providing their realtionship to partial quantum m-tuplet (or m-particle) distribution functions[4] (the generalization of the more commonly used partial quantum radial or pair distribution function). The main application emphasis of the method of this paper is to weakly degenerate, weakly interacting systems which is presented in section IV. Since perturbative methods suffice for investigation in this thermodynamic regime, our focus in this section is largely in the derivation of the second virial coefficient for quantum mixed systems. A valid expression is also given for PQNNPDF's in this section, in the region of medium/strong degeneracy, where particle interactions are however considered weak. This is especially of interest for the emerging field dealing with cold neutral atom systems trapped in magnetic/optical fields.

# II. Statistical Thermodynamic Formalism For Equilibrium Multi-component Quantum Materials

The translational coordinates in Eq. (1) may be ordered in the sense of one of the stated versions of the generalized ordering scheme given in Refs. 1, 2, and this should be evident in the way the domain of integration is specified. Writing the resulting generalized ordered translational coordinates as $X'_{k'_i i}$, the domain of integration of Eq. (1) may be specified such that the volume $v_i$ of the set of points traced out by $X'_{k'_i i}$ for the particle species $k_i'$ with label i has the property $0 \le v_i \le v_{i-1}$ (see Refs 1, 2). Eq. (1) may therefore be written as

$$Z(N_1,\cdots,N_n;V) = \sum_{k'_1,\ldots,k'_N} \int\cdots\int W_{N_1,\ldots,N_n} \prod_{i=1}^{N} dX'_{k'_i i} \quad\text{-----------------(5)}$$

where $(\prod_{i=1}^{n} \sigma_i^{N_i})\widetilde{Z}(N_1,\cdots,N_n;V) = Z(N_1,\cdots,N_n;V)$, and the sum is over different distinct combinations of allowed $k'_1,\cdots,k'_N$ values. (Remember exactly $N_j$ of the $k'_i$'s must assume the value j). This implies a total of $\dfrac{N!}{N_1!\cdots N_n!}$ terms for the above sum which is actually the number of distinct ways the particles may be relabeled or reassigned to the volumes $v_1, \ldots, v_N$.

These terms can be grouped as follows. The i-th group (i = 1, …, n) has terms for which the first labeled particle associated with volume $v_1$, is a particle of the i-th species (ie, $k'_i = i$). The i-th group therefore involves $\dfrac{(N-1)!}{N_1!\cdots N_{i-1}!(N_i-1)!N_{i+1}!\cdots N_n!}$ terms. Eq. 5 may therefore be rewritten as:

$$Z(N_1,N_2,\ldots,N_n,V) = \sum_{j=1}^{n} \int [\sum_{k'_2,\ldots,k'_N} \int\cdots\int W_{N_1,\ldots,N_n} \prod_{i=2}^{N} dX'_{k'_i i}] dX'_{j1} \quad\text{--------------(6)}$$

The first group (discussed above) involves the j = 1 summand of the outer sum of Eq. 6, the second group involves the j = 2 summand of the outer sum, etc. We may write:

$$W_{N_1,\ldots,N_n}(X'_{k'_1 1},\ldots,X'_{k'_N N}) = W_{N_1,\ldots,N_{i-1},N_i-1,N_{i+1},\ldots,N_n}(X'_{k'_2 2},\ldots,X'_{k'_N N})\left[\dfrac{W_{N_1,\ldots,N_n}(X'_{k'_1 1},\ldots,X'_{k'_N N})}{W_{N_1,\ldots,N_{i-1},N_i-1,N_{i+1},\ldots,N_n}(X'_{k'_2 2},\ldots,X'_{k'_N N})}\right]$$

$$\text{--------------(7)}$$

Taking $X'_{k'_1 1}$ as origin, and reordering the other generalized ordered coordinates according to distance from the origin, we may write $X_{k_1 1}$ as the coordinate (from among the remaining generalized ordered coordinates) which is nearest to the origin; $X_{k_2 2}$ the coordinate (from among the remaining generalized ordered coordinates) which is second nearest to the origin, etc. We may consider 2 clusters of coordinates; the first involving the coordinates $X'_{k'_1 1}, X_{k_1 1}, \cdots, X_{k_m m}$ and the second involving $X_{k_{m+1}(m+1)}, \cdots, X_{k_{N-1}(N-1)}$. For some finite m which is large enough (but not expected to be too large), the second mild statement of the product property of the Multi-component Slater Sum (given in section 1) allows us write the term in square brackets in Eq. (7) accurately as $\left[ \dfrac{W_{m_1,..,m_{i'-1},(m_{i'}+1),m_{i'+1},..,m_n}(X'_{k'_1 1}, X_{k_1 1},..., X_{k_m m})}{W_{m_1,...,m_n}(X_{k_1 1},..., X_{k_m m})} \right]$. (It is assumed $k_1' = i'$).

Accuracy in this case meaning that

$\left| w_{m_1,..,m_{i'-1},(m_{i'}+1),m_{i'+1},..,m_n} - w_{m_1,...,m_n} \right| \gg \left| f_{m_1,..,m_{i'-1},(m_{i'}+1),m_{i'+1},..,m_n} - f_{m_1,...,m_n} \right|$ or

$f_{m_1,..,m_{i'-1},(m_{i'}+1),m_{i'+1},..,m_n} - f_{m_1,...,m_n} \sim 0$; See section I. Note that while the first cluster is of finite size, the second cluster is of infinite size in the thermodynamic limit. Eq. (6) may therefore be written as:

$$Z(N_1, \cdots N_n, V) = \sum_{j=1}^{n} \int Z(N_1,..,N_{j-1}, N_j - 1, N_{j+1},.., N_n, v_1) \times$$

$$\times [\sum_{k'_2,...,k'_N} \int ... \int \left[ \dfrac{W_{m_1,..,m_{j-1},(m_j+1),m_{j+1},..,m_n}(X'_{j1}, X_{k_1 1},..., X_{k_m m})}{W_{m_1,...,m_n}(X_{k_1 1},..., X_{k_m m})} \right]$$

$$\times \left\{ W_{N_1,..,N_{j-1},N_j-1,N_{j+1},..,N_n}(X'_{k'_2 2},..., X'_{k'_N N}) \middle/ \sum_{k'_2,...,k'_N} \int ... \int W_{N_1,..,N_{j-1},N_j-1,N_{j+1},..,N_n}(Y'_{k'_2 2},..., Y'_{k'_N N}) \prod_{i=2}^{N} dY'_{k'_i i} \right\} \times$$

$$\times \prod_{l=2}^{N} dX'_{k'_l l} ] dX'_{j1}$$

-------------(8)

The volume within which the variables $X'_{k'_2 2}, \cdots, X'_{k'_N N}$ (as well as $Y'_{k'_2 2}, \cdots, Y'_{k'_N N}$ which are also generalized ordered) are restricted, being $v_1$. The volume $v_1$, is that traced out by the coordinate $X'_{j1}$ (see Refs 1 – 3). Also, we have that

$$Z(N_1, \cdots, N_{j-1}, N_j - 1, N_{j+1}, \cdots, N_n, v_1) = \sum_{k'_2,...,k'_N} \int \cdots \int W_{N_1,..,N_{j-1},N_j-1,N_{j+1},..,N_n}(X'_{k'_2 2},..., X'_{k'_N N}) \prod_{i=2}^{N} dX'_{k'_i i}$$

From Eq. (8), we may write:

$$Z(N_1,\cdots,N_n,V) = \sum_{j=1}^{n} \int Z(N_1,\cdots,N_{j-1},N_j-1,N_{j+1},\cdots,N_n,v_1) P_j(X'_{j1},v_1) dX'_{j1}$$

------------(9)

where:

$$P_j(X'_{j1},v_1) = \sum_{k'_2,\ldots,k'_N} \int \cdots \int \left[ \frac{W_{m_1,\ldots,m_{j-1},(m_j+1),m_{j+1},\ldots,m_n}(X'_{j1},X_{k_11},\ldots,X_{k_m m})}{W_{m_1,\ldots,m_n}(X_{k_11},\ldots,X_{k_m m})} \right] g^Q_{1,\ldots,N-1}(X'_{k'_2 2},\ldots,X'_{k'_N N}) \prod_{i=2}^{N} dX'_{k'_i i}$$

$(m \gg 1)$

----------(10)

$g^Q_{1,\ldots,N-1}(X'_{k'_2 2},\cdots,X'_{k'_N N})$ is the term in curly brackets in Eq. (8). As in the single component quantum case[3], in the condition m >> 1, it is tacitly assumed that "reasonably" accurate results can be obtained with $m \leq$ (a few tens). The coordinates $X_{k_i i}$ (i = 1, …, N – 1) re-ordered according to distance from the origin ($X'_{j1}$) may also clearly be said to be ordered according to one of the versions of the generalized ordering scheme. If for instance the first version of the generalized ordering scheme is employed and the coordinates $X_{k_i i}$ (i = 1, …, N – 1) are restricted within the volume $v_1$, then the complete coordinate set $X'_{j1}, X_{k_1 1},\cdots,X_{k_{N-1}(N-1)}$ may be said to constitute a set which is ordered according to the first version of the generalized ordering scheme. Hence this new set of coordinates may be used to replace the set $X'_{k'_1 1},\cdots,X'_{k'_N N}$ in the above equations. The function $g^Q_{1,\ldots,N-1}(X_{k_1 1},\cdots,X_{k_{N-1}(N-1)})$ may therefore be seen as the analog of the partial NNPDF for classical systems of Ref 1, and is referred to as the partial quantum nearest neighbor probability density function (PQNNPDF). Since $W_{N_1,\ldots,N_n}$ is proportional to an actual probability density function in coordinate space, PQNNPDF's are also actual probability density functions in coordinate space for quantum systems in the non-relativistic limit. As in the classical multi-component problem[1], $P_j(X'_{j1},v_1)$ is the average of the term $\left[ \frac{W_{m_1,\ldots,m_{j-1},(m_j+1),m_{j+1},\ldots,m_n}(X'_{j1},X_{k_1 1},\ldots,X_{k_m m})}{W_{m_1,\ldots,m_n}(X_{k_1 1},\ldots,X_{k_m m})} \right]$ over the phase space $Z(N_1,\cdots,N_{j-1},N_j-1,N_{j+1},\cdots,N_n,v_1)$. Now, $P_j(X'_{j1},v_1)$ has a functional form which depends on $v_1$, the coordinate $X'_{j1}$, and the species type (which in this case is j) whose particle is said to be the first labeled particle associated with volume $v_1$. Also, the quantity $\left[ \frac{W_{m_1,\ldots,m_{j-1},(m_j+1),m_{j+1},\ldots,m_n}(X'_{j1},X_{k_1 1},\ldots,X_{k_m m})}{W_{m_1,\ldots,m_n}(X_{k_1 1},\ldots,X_{k_m m})} \right]$ depends on $X'_{j1}$, the species type indicated as j, and the coordinates of only the m-nearest neighbors of the origin ($X'_{j1}$). Hence $P_j(X'_{j1},v_1)$ may be rewritten in terms of PQNNPDF's as

$$P_j(X'_{j1}, v_1) = \sum_{(l)} \int \cdots \int \left[ \frac{W_{m_1,\ldots,m_{j-1},(m_j+1),m_{j+1},\ldots,m_n}(X'_{j1}, X_{k_1 1},\ldots, X_{k_m m})}{W_{m_1,\ldots,m_n}(X_{k_1 1},\ldots, X_{k_m m})} \right] g^{Q(l)}_{1,\ldots,m}(X_{k_1 1},\ldots, X_{k_m m}) \prod_{i=1}^{m} dX_{k_i i}$$

$(m \gg 1)$

-----------(11)

$g^{Q(l)}_{1,\ldots,m}(X_{k_1 1},\cdots, X_{k_m m})$ is referred to as the "general multi-component point process" (GMPP) PQNNPDF for m nearest neighbors (with origin at a point on the boundary of volume $v_1$). This is the analog of the GMPP PNNPDF for classical systems[1]. The index (*l*) indicates a specific choice of values for $k_1,\cdots,k_m$ of which there are $n^m$ of such choices. (It is assumed m << $N_i$ for i = 1, … , n, and so the choice of species for the first, second, etc. nearest neighbors may each be made in n ways). Generalizations to the "ordinary multi-component point process" (OMPP)[1] where a particle is situated at the origin, as well as cases where the origin is situated in the middle of the volume V are rather straightforward to also derive. It should be noted that as in the classical case for PNNPDF's[1] we also have that PQNNPDF's do not normalize by mere integration alone. Clearly, in Eq. (11), if we set the quantity $\left[ \dfrac{W_{m_1,\ldots,m_{j-1},(m_j+1),m_{j+1},\ldots,m_n}(X'_{j1}, X_{k_1 1},\ldots, X_{k_m m})}{W_{m_1,\ldots,m_n}(X_{k_1 1},\ldots, X_{k_m m})} \right]$ to some constant (such as unity), the average of the quantity (which is $P_j$) must equal the constant, and thus this requires the sum of eq. (11) to add up to the constant value, and this easily indicates how PQNNPDF's must normalize. PQNNPDF's are further discussed in the next section. Because $W_{m_1,\ldots,m_n}$ features in the function $P_j(X'_{j1}, v_1)$ as well as in $g^{Q(l)}_{1,\ldots,m}(X_{k_1 1},\cdots, X_{k_m m})$ (see next section), it is thus necessary to have an accurate and efficient way of computing $W_{m_1,\ldots,m_n}$ for various nearest neighbor configurations. In the literature,[4,5] extreme difficulties encountered in obtaining reasonable $W_{m_1,\ldots,m_n}$ solutions and related results, has given rise to a variety of model methods in condensed matter physics. Limitations and general inability to assess the efficacy of these methods, require that the methods constantly undergo revision and change, with newer schemes (only very few of which endure for long) introduced every so often. Despite these difficulties, it is expected that it is in cases of small m values we may hope to eventually achieve considerable progress in the bid to obtain accurate results for $W_{m_1,\ldots,m_n}$. (see Ref 3 for additional discussion). Now, $P_j(X'_{j1}, v_1)$ may generally vary as $X'_{j1}$ varies for fixed $v_1$. Hence we may replace $P_j$ by $<P_j>$ as in the classical case[1], and thus rewrite Eq. (9) as:

$$Z(N_1,\cdots,N_n,V) = \sum_{i=1}^{n} \int_0^V Z(N_1,\cdots,N_{i-1},N_i-1,N_{i+1},\cdots,N_n,v_1) <P_i> dv_1$$

-----------(12)

As in Ref 1, we expect $<P_j> \approx P_j$ almost always in many cases of interest. $W_{m_1,\ldots,m_n}$ is dimensionless, hence we may write

$$Z(N_1,\cdots,N_n,V) = \frac{1}{N_1!\cdots N_n!}\prod_{i=1}^{n}(\varepsilon V)^{N_i} \qquad \text{-----------(13)}$$

where ε is a dimensionless quantity which is some function of $N_1, \ldots, N_n$, V. As in the single component quantum case[3], ε is not expected to tend to unity in general in the limit of non-interaction (unlike the classical multi-component poisson point process) due to quantum effects. In the limit of very low densities however, ε is expected to tend to unity. Employing Eq. (13) in Eq. (12), and differentiating Eq. (12) with respect to V yields:

$$N\varepsilon^{N-1}(N_1,\cdots,N_n,V)\left[\varepsilon(N_1,\cdots,N_n,V) + V\frac{\partial \varepsilon(N_1,\cdots,N_n,V)}{\partial V}\right]$$
$$= \sum_{i=1}^{n} <P_i> N_i \varepsilon^{N-1}(N_1,..,N_{i-1},N_i-1,N_{i+1},..,N_n,V)$$
$$\text{-----------(14)}$$

Eq. (14) is valid for all $N_1, \ldots, N_n$, V values. Hence the developments in this paper is said to be applicable for all system scales; and Eq. (14) may be re-expressed for different limiting regimes corresponding to different system scales which may be variously classified, for instance as mesoscale, microscale, macroscale, etc. For example, in the thermodynamic limit, considered as the macroscale regime, it is assumed $\varepsilon(N_1,\cdots,N_n,V)$ may be rewritten as a function of only the partial densities $\rho_1, \ldots, \rho_n$. Hence proceeding as in Ref 1, Eq. (14) becomes:

$$\varepsilon\left(1 - \frac{\rho_1}{\varepsilon}\frac{\partial \varepsilon}{\partial \rho_1} - \cdots - \frac{\rho_n}{\varepsilon}\frac{\partial \varepsilon}{\partial \rho_n}\right) = \sum_{i=1}^{n}\alpha_i <P_i> \exp\left(-\frac{\rho}{\varepsilon}\frac{\partial \varepsilon}{\partial \rho_i}\right) \qquad \text{----------(15)}$$

Eq. (15) is then readily solved via an iterative scheme for ε employing methods outlined in Ref 1, 2. In the next section, we show that PQNNPDF's depend on $\varepsilon$; hence as $\varepsilon$ gets more accurately determined by the iterative process, PQNNPDF's also get more accurately evaluated. At the end of the iterative process, the free energy (obtained from $\varepsilon$) or PQNNPDF's may be employed to accurately compute a variety of system properties including chemical potentials, equation of state, etc. much like those for classical multi-component systems[1]. (We note however that since much calculations will be required to carry out the above numerical scheme, we have focused our main application at the present time to a thermodynamic regime where Eq. 15 may be more simply solved in one parse of the iteration process – see section IV).

    General expressions can be given for various properties of quantum mixed systems in terms of the parameter $\varepsilon$. For instance, the chemical potential of species i may be written as:

$$\mu_i = \left.\frac{\partial F}{\partial N_i}\right)_{T,V,N_j(j\neq i)} = F(T,V,N_1,..,N_{i-1},N_i+1,N_{i+1},..,N_n) - F(T,V,N_1,...,N_n)$$

where $F = -kT \ln \tilde{Z}$ is the Helmholtz free energy. Hence using Eq. (13) we have:

$$\mu_i = -kT\left[-\ln \sigma_i + \ln V - \ln(N_i+1) + \ln(\tilde{\varepsilon}^{N+1}) - \ln(\varepsilon^N)\right]$$

where $\tilde{\varepsilon} = \varepsilon(N_1,..,N_{i-1},N_i+1,N_{i+1},..,N_n,V)$. We consider only the macro-scale limit where the following asymptotic expressions are valid:

$$\tilde{\varepsilon} \approx \varepsilon + \frac{\partial \varepsilon}{\partial N_i}$$

$$\ln\left(1 + \frac{1}{\varepsilon}\frac{\partial \varepsilon}{\partial N_i}\right) \approx \frac{1}{\varepsilon}\frac{\partial \varepsilon}{\partial N_i}$$

(It is assumed $\frac{1}{\varepsilon}\frac{\partial \varepsilon}{\partial N_i} \ll 1$, and $\varepsilon$ can be written as a function of $\rho_1, \ldots, \rho_n$ in place of $N_1, \ldots, N_n, V$). Hence in the macro-scale limit we have:

$$\mu_i = -kT\left[\ln\left(\frac{\varepsilon}{\rho_i}\right) + \frac{\rho}{\varepsilon}\frac{\partial \varepsilon}{\partial \rho_i} - \ln \sigma_i\right] \quad\text{-----------(16)}$$

Following developments in Ref. 1, the quantum mixed law of mass action is

$$\prod_{i=1}^{n}\rho_i^{\nu_i} = \prod_{i=1}^{n}\left(\varepsilon e^{-\phi_i+1}\right)^{\nu_i}\prod_{i=1}^{n}\sigma_i^{-\nu_i} \quad\text{-----------(17)}$$

The $\nu_i$'s are stoichiometric coefficients in the associated balanced chemical equation. $\phi_i = p_i/\rho_i kT$, where $p_i$ may be regarded as the partial pressure of the i-th species which is written as: $p_i = kT\rho_i\left(1 - \frac{\rho}{\varepsilon}\frac{\partial \varepsilon}{\partial \rho_i}\right)$.

The classical limiting behavior for $W_{m_1,\ldots,m_n}$ is proved the same way as for the single component quantum case[3]. Hence we may readily state that if the "effective" pair-wise inter-particle interaction potential is largely weak and slowly varying (so that at worst, its characteristic distance is $\gg \lambda_i$ for i = 1, ..., n), and if the average distance between the coordinates of the m-nearest neighbors ($X_{k_11},\cdots,X_{k_mm}$) is also $\gg \lambda_i$ (for i = 1, ..., n), we have that $W_{m_1,\ldots,m_n}$ approximates to:

$$\prod_{i=1}^{n}\left(\frac{\lambda_i}{h}\right)^{3m_i}\int\cdots\int e^{-\beta \tilde{H}_{m_1,\ldots,m_n}}dp_{k_11}\cdots dp_{k_mm} = e^{-\beta U_{m_1,\ldots,m_n}}$$

$\tilde{H}_{m_1,\ldots,m_n}$ is the classical Hamiltonian for $m_i$ species i particles for i ranging from 1 to n. $p_{k_i i}$ is the momentum conjugate to $X_{k_i i}$. ($m_i$ of the species indices $k_1, \ldots, k_m$ have the value i for i = 1, ..., n; and $m_1 + \ldots + m_n = m$). $U_{m_1,\ldots,m_n}$ is the potential energy of the m-nearest neighbors (in the GMPP process) of the origin (where effect from the remaining N – m particles of the system is not included). Hence the quantity

$$\left[ \frac{W_{m_1,\ldots,m_{j-1},(m_j+1),m_{j+1},\ldots,m_n}(X'_{j1}, X_{k_1 1},\ldots, X_{k_m m})}{W_{m_1,\ldots,m_n}(X_{k_1 1},\ldots, X_{k_m m})} \right]$$

reduces to the quantity $e^{-\beta E_{j2}}$ of Ref. 1, and the quantum calculations for $Z(N_1,\cdots,N_n,V)$ is said to coincide with those of the classical case in the low density and/or high temperature limits where PQNNPDF's are expected to give much more weight to configurations in which coordinates are far apart compared to $\lambda_i$.

# III. PQNNPDF's For Microstructure in Quantum Mixed Systems

In the previous section, PQNNPDF's were introduced and shown to be the counterparts of classical case PNNPDF's for describing structure in quantum systems. PQNNPDF's are now developed further in this section. Defining an origin somewhere in the middle of volume V, coordinates of particles are assumed to be ordered according to their radial distance from the origin. ie. $r_{k_{i+1}(i+1)} \geq r_{k_i i}$ for $i = 1, \cdots, N-1$ ($r_{k_i i}$ is the radial part of $X_{k_i i}$). This implies the coordinates are generalized ordered. We can cast PQNNPDF's as thermal averages of appropriately formulated particle number density operators. Writing the variable of integration over generalized ordered coordinates as $\hat{X}_{k_i i}$, we write the particle number density operator for the i-th nearest neighbor (species $k_i$) of the origin as $\delta(\hat{X}_{k_i i} - X_{k_i i})$, where δ is the dirac delta function. The equilibrium thermal density operator for the system is $\hat{\rho} = \exp(-\beta H_{N_1, \cdots, N_n})$. Hence we may write the GMPP PQNNPDF for all N coordinates for instance as:

$$g_{1,\cdots,N}^{QG}(X_{k_1 1}, \cdots, X_{k_N N}) = \langle \delta(\hat{X}_{k_1 1} - X_{k_1 1}) \cdots \delta(\hat{X}_{k_N N} - X_{k_N N}) \rangle$$

$$= \tilde{Z}^{-1}(N_1, \cdots, N_n, V) Tr\left(\hat{\rho} \prod_{i=1}^{N} \delta(\hat{X}_{k_i i} - X_{k_i i})\right)$$

$$= Z^{-1} \int \cdots \int W_{N_1, \cdots, N_n}(\hat{X}_{k_1 1}, \cdots, \hat{X}_{k_N N}) \prod_{i=1}^{N} \left[d\hat{X}_{k_i i} \delta(\hat{X}_{k_i i} - X_{k_i i})\right]$$

$$= Z^{-1} W_{N_1, \cdots, N_n}(X_{k_1 1}, \cdots, X_{k_N N})$$

Writing $W_{N_1, \cdots, N_n}$ as $W_{N_1 - m_1, \cdots, N_n - m_n} W_{m_1, \cdots, m_n} F_{m_1, \cdots, m_n}$, we can write the PQNNPDF for m nearest neighbors ($m_i$ of which are of species i, and $m_1 + \cdots + m_n = m$) in the multi-component point process as a thermal average of the operator $\delta(\hat{X}_{k_1 1} - X_{k_1 1}) \cdots \delta(\hat{X}_{k_m m} - X_{k_m m})$ as:

$$g_{1,\cdots,m}^{QG}(X_{k_1 1}, \cdots, X_{k_m m}) = \exp\left(\frac{A}{kT}\right) Tr\left(\hat{\rho} \prod_{i=1}^{m} \delta(\hat{X}_{k_i i} - X_{k_i i})\right)$$

$$= \left(\prod_{i=1}^{n} \sigma_i^{N_i}\right)^{-1} \exp\left(\frac{A}{kT}\right) W_{m_1, \cdots, m_n} \int \cdots \int W_{N_1 - m_1, \cdots, N_n - m_n} F_{m_1, \cdots, m_n} \prod_{i=m+1}^{N} dX_{k_i i}$$

----------(18)

The coordinates $X_{k_i i}$ ($i = m+1, \cdots, N$) in Eq. (18) are restricted within volume $\hat{V} = V - \frac{4}{3}\pi r_{k_m m}^3$ which has a different "shape" than that of volume V. Following the method outlined in Edgal[8] (1991), Eq. (18) is manipulated to arrive at:

$$g_{1,\ldots,m}^{QG}\left(X_{k_1 1}, \cdots, X_{k_m m}\right) = \left[\left(\prod_{i=1}^{n}\sigma_i^{N_i}\right)^{-1}\exp\left(\frac{A}{kT}\right)Z(N_1 - m_1, \cdots, N_n - m_n, V)\right] \times$$

$$\times \left[\frac{{}^{\hat{V}}S_{N_1-m_1,\ldots,N_n-m_n}\langle F_{m_1,\ldots,m_n}\rangle Z(N_1 - m_1, \cdots, N_n - m_n, \hat{V})}{Z(N_1 - m_1, \cdots, N_n - m_n, V)}\right]\left[W_{m_1,\ldots,m_n}\right]$$

(19)

${}^{\hat{V}}S_{N_1-m_1,\ldots,N_n-m_n}$ is a shape effect factor which accounts for the difference in shape between the volume $\hat{V}$ which has a "void" of size $\frac{4}{3}\pi r_{k_m m}^3$ located within it, and some volume of "standard" shape (without a void) of the same size. (Note that the shape of volume V actually defines the standard shape – see Edgal[8] (1991) for a detailed discussion). $\langle F_{m_1,\ldots,m_n}\rangle$ is an average taken over the phase space of size $Z_s(N_1 - m_1, \cdots, N_n - m_n, \hat{V}) = {}^{\hat{V}}S_{N_1-m_1,\ldots,N_n-m_n}Z(N_1 - m_1, \cdots, N_n - m_n, \hat{V})$. (The quantity $Z(N_1 - m_1, \cdots, N_n - m_n, \hat{V})$ is the partition function apart from the factor of $\left(\prod_{i=1}^{n}\sigma_i^{N_i-m_i}\right)$, evaluated using the volume of standard shape of size $\hat{V}$). ${}^{\hat{V}}S_{N_1-m_1,\ldots,N_n-m_n}$ and $\langle F_{m_1,\ldots,m_n}\rangle$ provide shape and surface effects respectively, as discussed in Edgal[8] (1991). Expressing $Z(N_1 - m_1, \cdots, N_n - m_n, V)$ in terms of $\varepsilon$ and noting $|w_{m_1,\ldots,m_n}|, |w_{N_1-m_1,\ldots,N_n-m_n}| \gg |f_{m_1,\ldots,m_n}|$ for m sufficiently large (cf section 1), we may argue as in Edgal[8] (1991) that because surface and shape effects present themselves in "reduced" forms, they may be ignored (similar to small computational errors such as round off errors) once we have chosen m to be as large as some value that is "small" (probably a few tens or so). Eq. (19) can be considered in various asymptotic limits (regarded as micro-scale, meso-scale, macro-scale, etc). In particular, in the macro-scale or thermodynamic limit, Eq. (19) may be rewritten as:

$$g_{1,\ldots,m}^{QG}\left(X_{k_1 1}, \cdots, X_{k_m m}\right) = h_{k_1 1,\ldots,k_m m}\exp\left[-\frac{4}{3}\pi r_{k_m m}^3 \rho\left(1 - \frac{\rho_1}{\varepsilon}\frac{\partial\varepsilon}{\partial\rho_1} - \cdots - \frac{\rho_n}{\varepsilon}\frac{\partial\varepsilon}{\partial\rho_n}\right)\right]W_{m_1,\ldots,m_n}$$

$$= h_{k_1 1,\ldots,k_m m}\exp\left[-\frac{4}{3}\pi r_{k_m m}^3 \frac{p}{kT}\right]\left[W_{m_1,\ldots,m_n}\left(X_{k_1 1}, \cdots, X_{k_m m}\right)\right]$$

$(m \gg 1)$

(20)

$h_{k_1 1, \ldots, k_m m}$ is called a partial normalization constant for reasons to be discussed soon. p is the system's pressure. Use has been made of the fact that we can write the equation of state (as in the classical multi-component system)[1] in the thermodynamic limit as:

$$\phi = \frac{p}{kT} = \rho \left( 1 - \frac{\rho_1}{\varepsilon} \frac{\partial \varepsilon}{\partial \rho_1} - \cdots - \frac{\rho_n}{\varepsilon} \frac{\partial \varepsilon}{\partial \rho_n} \right) \qquad (21)$$

By arguments given earlier, in respect of Eq. (11) in section II, the PQNNPDF of Eq. (20) does not normalize by mere integration as in the single component case, but must normalize as follows:

$$\sum_{(l)} \int \cdots \int g_{1,\ldots,m}^{QG(l)}\left(X_{k_1 1}, \cdots, X_{k_m m}\right) \prod_{i=1}^{m} dX_{k_i i} = 1 \qquad (22)$$

$g_{1,\ldots,m}^{QG}\left(X_{k_1 1}, \cdots, X_{k_m m}\right)$ is simply rewritten as $g_{1,\ldots,m}^{QG(l)}\left(X_{k_1 1}, \cdots, X_{k_m m}\right)$ in Eq. (22), and the index $(l)$ is used to indicate a specific choice of values of $k_1, \cdots, k_m$, implying the sum involves $n^m$ terms. Hence there are $n^m$ of the constants, $h_{k_1 1, \ldots, k_m m}$, that must feature in the normalization of PQNNPDF's. Thus the constants $h_{k_1 1, \ldots, k_m m}$ are rightly termed partial normalization constants while $g_{1,\ldots,m}^{QG(l)}\left(X_{k_1 1}, \cdots, X_{k_m m}\right)$ is rightly seen as a partial probability density function. Eq. (20) is "exact", and it rigorously describes structure in quantum systems. In the OMPP case, PQNNPDF's may be rewritten as $g_{1,\ldots,m}^{Qi}\left(X_{k_1 1}, \cdots, X_{k_m m}\right)$ or $g_{1,\ldots,m}^{Qi(l)}\left(X_{k_1 1}, \cdots, X_{k_m m}\right)$ for the case where the particle at the origin is of species i. Also, $N_i$ is written as $N_i - 1$, and the multi-component Slater Sum, $W_{m_1,\ldots,m_n}(X_{k_1 1}, \ldots, X_{k_m m})$ is changed to $W_{m_1,\ldots,m_{i-1},(m_i+1),m_{i+1},\ldots,m_n}(X_{i0}, X_{k_1 1}, \ldots, X_{k_m m})$. The coordinate $X_{i0}$ indicates the particle (species i) at the origin is also considered in evaluating the Slater Sum. The free energy A in Eq. (18) becomes that for $N_1$ (species 1) particles,..., $N_{i-1}$ (species i – 1) particles, $N_i - 1$ (species i) particles, $N_{i+1}$ (species i + 1) particles,..., $N_n$ (species n) particles in a space of volume V, causing some modification of the normalization factor $h_{k_1 1, \ldots, k_m m}$. For a GMPP process for instance in which the origin is located at a point on the boundary surface of volume V, it is easy to see that Eq. (20) becomes

$$g_{1,\ldots,m}^{Q}\left(X_{k_1 1}, \cdots, X_{k_m m}\right) = h_{k_1 1, \ldots, k_m m} \exp\left[-\frac{2}{3}\pi r_{k_m m}^3 \frac{p}{kT}\right]\left[W_{m_1,\ldots,m_n}\left(X_{k_1 1}, \cdots, X_{k_m m}\right)\right]$$
$$(m \gg 1)$$
$$\qquad (23)$$

It is assumed that the boundary surface of volume V is locally flat everywhere; hence the factor of 4/3 in Eq. (20) is simply replaced by 2/3 in the macro-scale limit (c.f. Ref. 1). Eq. (23) is what is required to be used for the PQNNPDF in Eq. (11) where the effect of the particle located at the origin $X'_{k'_i i}$ (which is a point on the boundary surface of volume $v_1$) is not taken into account in the evaluation of $W_{m_1,...,m_n}(X_{k_1 1},...,X_{k_m m})$ (being of course a GMPP distribution). The way that terms in Eq. (23) or Eq. (20) compete in different temperature and density regimes, to provide various features of nearest neighbor configurations, can be studied employing such terminology as "push" effect, as was done for the single component quantum case.[3]

As discussed for the classical multi-component case[8], an added level of difficulty is encountered in handling quantum mixed systems since several partial normalization constants are needed to normalize PQNNPDF's. However, also as discussed for the classical multi-component case, this difficulty is obviated by simulating only a single but fairly large multi-component system where m is not only larger than n, but is large enough for it to be exactly or approximately true that the number of particles (species i) involved in the simulation is equal to $\alpha_i m \gg 1$ (for $i = 1,\cdots,n$). Since distributions in such a large system is sharply peaked, the region of phase space ignored by not considering the variability in the number of particles of given species in volume $v_m$ is sufficiently small and safely ignored. As the large system is therefore made to perform a "random walk" through phase space (such as by the Metropolis algorithm), the ordering of particles (by species) according to distance from the origin, will change several times with appropriate "weighting". This amounts to considering simultaneously, the relevant set of PQNNPDF's, which constitutes a relatively "small" set, with each PQNNPDF having the same set of $m_i$ values but different partial normalization constants. By integrating PQNNDPF's for such large systems to obtain PQNNPDF's of smaller m values, we may then expect to obtain accurate expressions for accurate study of Eq. (20) (including studies of local partial density fluctuations for different species).

Partial m-tuplet distribution functions (variously called partial m-particle distribution, partial correlation function, etc) can also be formulated as has been done for classical multi-component systems[8]. Since these distributions are presently much more used than the emergent distributions involving PQNNPDFs for describing microstructure of arbitrary material systems (from crystals to complex highly disordered systems)[4,6], a brief treatise of the distributions and their relationship to PQNNPDF's will now be given.

Writing particle coordinates that are **not** generalized ordered as in section I as $X''_{k''_1 1},\cdots,X''_{k''_N N}$, we write the N-tuplet distribution function for all N particles in a quantum mixed system as the following thermal average:

$$F_{N_1,...,N_n}\left(X''_{11},\cdots,X''_{1N_1};\cdots\cdots;X''_{n1},\cdots,X''_{nN_n}\right) = \left\langle \prod_{j=1}^{n}\left(\frac{\delta(\hat{X}''_{j1}-X''_{j1})\cdots\delta(\hat{X}''_{jN_j}-X''_{jN_j})}{N_j!}\right)\right\rangle_{NGO}$$

$$= \left(\prod_{i=1}^{n}\sigma_i^{N_i}\right)^{-1} \frac{\exp(A/kT)}{N_1!\cdots N_n!} W_{N_1,...,N_n}\left(X''_{11},\cdots,X''_{1N_1};\cdots\cdots;X''_{n1},\cdots,X''_{nN_n}\right)$$

(24)

The averaging process under the condition that coordinates are not generalized ordered implies particle configurations are considered distinct when coordinates of identical particles are merely permuted. Hence identical particles are treated as distinguishable in the averaging process (but definitely not from a quantum point of view); and since the free energy is computed assuming identical particles are indistinguishable, this is why the factor of $\frac{1}{N_1! \cdots N_n!}$ has been introduced. To distinguish this averaging process from the one in which particle coordinates are generalized ordered, the subscript index "NGO" has been used. The $F_{N_1,\ldots,N_n}$ distribution is vanishingly small in the thermodynamic limit, and so it is usually preferable to work with the distribution $f_{N_1,\ldots,N_n}(X''_{11},\ldots\ldots,X''_{nN_n})$ defined by

$$F_{N_1,\ldots,N_n}(X''_{11},\ldots\ldots,X''_{nN_n}) = \frac{1}{V^N} f_{N_1,\ldots,N_n}(X''_{11},\ldots\ldots,X''_{nN_n}) \qquad (25)$$

Reduced partial m-tuplet distribution function where identical particles are distinguishable can be defined as the thermal average (with the subscript index NGO):

$$\frac{1}{V^m} f_{m_1,\ldots,m_n}(X''_{11},\cdots,X''_{1m_1};\ldots\ldots;X''_{n1},\cdots,X''_{nm_n}) = \left\langle \prod_{j=1}^{n} \left( \frac{\delta(\hat{X}''_{j1} - X''_{j1})\cdots\delta(\hat{X}''_{jm_j} - X''_{jm_j})}{N_j!} \right) \right\rangle_{NGO}$$
(26)

(where $m_1 + \cdots + m_n = m$). For any j for which $m_j = 0$, the numerator of the argument of the product in Eq. (26) is unity. Comparing Eqs. 26 and 24, and using Eq. (25), we find that we may write:

$$\frac{1}{V^N} \int_V \cdots \int_V f_{N_1,\ldots,N_n}(X''_{11},\ldots,X''_{nN_n}) \prod_{t_1=m_1+1}^{N_1} dX''_{1t_1} \cdots \prod_{t_n=m_n+1}^{N_n} dX''_{nt_n} = \frac{1}{V^m} f_{m_1,\ldots,m_n}(X''_{11},\ldots,X''_{1m_1};\ldots;X''_{n1},\ldots,X''_{nm_n})$$

Hence we may more generally write:

$$\frac{1}{V^{s-m}} \int_V \cdots \int_V f_{s_1,\ldots,s_n}(X''_{11},\ldots,X''_{1s_1};\ldots;X''_{n1},\ldots,X''_{ns_n}) \prod_{t_1=m_1+1}^{s_1} dX''_{1t_1} \cdots \prod_{t_n=m_n+1}^{s_n} dX''_{nt_n}$$
$$= f_{m_1,\ldots,m_n}(X''_{11},\ldots,X''_{1m_1};\ldots;X''_{n1},\ldots,X''_{nm_n}) \qquad (27)$$

The normalization of partial m-tuplet distribution function follows as:

$$\frac{1}{V^m} \int_V \cdots \int_V f_{m_1,\ldots,m_n}(X''_{11},\ldots,X''_{1m_1};\ldots;X''_{n1},\ldots,X''_{nm_n}) \prod_{t_1=1}^{m_1} dX''_{1t_1} \cdots \prod_{t_n=1}^{m_n} dX''_{nt_n} = 1 \qquad (28)$$

Conditional partial m-tuplet distribution function (under the condition of distinguishability of identical particles) can be defined for the event that some specific $m_1$ particles of species 1 are located at $X''_{11},\cdots,X''_{1m_1}$; some specific $m_2$ particles of species 2 are located at $X''_{2(m_1+1)},\cdots,X''_{2(m_1+m_2)}$; ……; some specific $m_n$ particles of species n are located at $X''_{n(m_1+\ldots+m_{n-1}+1)}, \cdots, X''_{n(m_1+\ldots+m_n)}$; given that some other specific $p_1$ particles of species 1 are located at $X''_{1(m_1+\ldots+m_n+1)}, \cdots, X''_{1(m_1+\ldots+m_n+p_1)}$; some other specific $p_2$ particles of species 2 are located at $X''_{2(m_1+\ldots+m_n+p_1+1)}, \cdots, X''_{2(m_1+\ldots+m_n+p_1+p_2)}$; ……; some other specific $p_n$ particles of species n are located at $X''_{n(m_1+\ldots+m_n+p_1+\ldots+p_{n-1}+1)}, \cdots, X''_{n(m_1+\ldots+m_n+p_1+\ldots+p_n)}$ is given as:

$$f_{m_1,\ldots,m_n}\left(X''_{k''_1 1},\ldots, X''_{k''_m m} \,\Big|\, X''_{k''_{m+1}(m+1)},\ldots, X''_{k''_{m+p}(m+p)}\right) = \frac{f_{(m_1+p_1),\ldots,(m_n+p_n)}\left(X''_{k''_1 1},\ldots, X''_{k''_{m+p}(m+p)}\right)}{f_{p_1,\ldots,p_n}(X''_{k''_{m+1}(m+1)},\ldots, X''_{k''_{m+p}(m+p)})} \tag{29}$$

(where $p_1 + \cdots + p_n = p$). This conditional m-tuplet distribution function normalizes as usual as:

$$\int_V \cdots \int_V \frac{1}{V^m} f_{m_1,\ldots,m_n}\left(X''_{k''_1 1},\ldots, X''_{k''_m m} \,\Big|\, X''_{k''_{m+1}(m+1)},\ldots, X''_{k''_{m+p}(m+p)}\right) dX''_{k''_1 1}\ldots dX''_{k''_m m} = 1 \tag{30}$$

For the more commonly used case where identical particles are treated as **indistinguishable** while coordinates are not generalized ordered, the thermal average employed is still the average with the subscript index NGO. However, the operator to be thermally averaged (with the NGO subscript index) for the reduced partial m-tuplet distribution function (now written as $\frac{1}{V^m}\tilde{f}_{m_1,\ldots,m_n}\left(X''_{11},\ldots, X''_{1m_1}; \ldots; X''_{n1},\ldots, X''_{nm_n}\right)$ or more simply as $\hat{f}_{m_1,\ldots,m_n}\left(X''_{11},\ldots, X''_{1m_1}; \ldots; X''_{n1},\ldots, X''_{nm_n}\right)$) may be written for instance for the single component case (n = 1) with m = 2 (for the 2-body distribution function) as $\sum_{\substack{t_1,t_2 \\ (t_1 \neq t_2)}} \frac{1}{N!} \delta(\hat{X}''_{1t_1} - X''_{11}) \delta(\hat{X}''_{1t_2} - X''_{12})$. This operator is more generally written for arbitrary m, n as:

$$\sum_{\substack{t_1,\ldots,t_{m_1} \\ (t_1 \neq \ldots \neq t_{m_1})}} \sum_{\substack{t_{m_1+1},\ldots,t_{m_1+m_2} \\ (t_{m_1+1} \neq \ldots \neq t_{m_1+m_2})}} \cdots \sum_{\substack{t_{m_1+\ldots+m_{n-1}+1},\ldots,t_{m_1+\ldots+m_n} \\ (t_{m_1+\ldots+m_{n-1}+1} \neq \ldots \neq t_{m_1+\ldots+m_n})}} \left[ \frac{\prod_{i=1}^{m_1} \delta(\hat{X}''_{1t_i} - X''_{1i}) \prod_{j=1}^{m_2} \delta(\hat{X}''_{2t_{m_1+j}} - X''_{2(m_1+j)}) \cdots \prod_{l=1}^{m_n} \delta(\hat{X}''_{nt_{m_1+\ldots+m_{n-1}+l}} - X''_{n(m_1+\ldots+m_{n-1}+l)})}{N_1!\ldots N_n!} \right]$$

The indices $(t_1,\ldots, t_{m_1})$ in the first multiple sum form an $m_1$-tuple; and the total number of $m_1$-tuples in the first multiple sum is determined as the number of distinct ways of choosing $m_1$ particles of species 1 from $N_1$ identical particles, multiplied by the number of ways of placing the $m_1$ chosen particles into locations $X''_{11}, \cdots, X''_{1m_1}$. This number is $\frac{N_1!}{(N_1-m_1)!m_1!} \times m_1! = \frac{N_1!}{(N_1-m_1)!}$. The total number of $m_2$-tuples in the second multiple sum is similarly determined as $\frac{N_2!}{(N_2 - m_2)!}$; etc. Hence the total number of terms in the above multiple sum is $\frac{N_1! \cdots N_n!}{(N_1 - m_1)! \cdots (N_n - m_n)!}$. (Note that it is assumed m << $N_i$. If m were $\gtrsim N_i$, it would generally be difficult to formulate the number of terms in the multiple sum. A careful observation reveals that each term of the multiple sum yields the same thermal average. Hence the operator to be thermally averaged (with the NGO subscript) for the reduced partial m-tuplet distribution function (where identical particles are considered indistinguishable) can be more simply written as:

$$\frac{N_1! \cdots N_n!}{(N_1 - m_1)! \cdots (N_n - m_n)!} \prod_{j=1}^{n} \left( \frac{\delta(\hat{X}''_{j1} - X''_{j1}) \cdots \delta(\hat{X}''_{jm_j} - X''_{jm_j})}{N_j!} \right)$$

Hence from Eq. (26), we have that:

$$\hat{f}_{m_1,\ldots,m_n} = \frac{N_1! \ldots N_n!}{(N_1 - m_1)! \ldots (N_n - m_n)!} \left( \frac{1}{V^m} f_{m_1,\ldots,m_n} \right) \tag{31}$$

The normalization of the new reduced partial m-tuplet distribution function where identical particles are treated as indistinguishable is:

$$\int_V \cdots \int_V \hat{f}_{m_1,\ldots,m_n} \left( X''_{k''_1 1}, \ldots, X''_{k''_m m} \right) dX''_{k''_1 1} \ldots dX''_{k''_m m} = \frac{N_1! \cdots N_n!}{(N_1 - m_1)! \cdots (N_n - m_n)!} \tag{32}$$

Conditional partial m-tuplet distribution function can be defined under the condition of indistinguishability of identical particles similar to Eq. (29) as:

$$\hat{f}_{m_1,\ldots,m_n} \left( X''_{k''_1 1}, \ldots, X''_{k''_m m} \middle| X''_{k''_{m+1}(m+1)}, \ldots, X''_{k''_{m+p}(m+p)} \right) = \frac{\hat{f}_{(m_1+p_1),\ldots,(m_n+p_n)} \left( X''_{k''_1 1}, \ldots, X''_{k''_{m+p}(m+p)} \right)}{\hat{f}_{p_1,\ldots,p_n} (X''_{k''_{m+1}(m+1)}, \ldots, X''_{k''_{m+p}(m+p)})} \tag{33}$$

Using Eq. (31), we have that this conditional partial m-tuplet distribution function normalizes as:

$$\int_V \cdots \int_V \hat{f}_{m_1,\ldots,m_n} \left( X''_{k''_1 1}, \ldots, X''_{k''_m m} \middle| X''_{k''_{m+1}(m+1)}, \ldots, X''_{k''_{m+p}(m+p)} \right) dX''_{k''_1 1} \cdots dX''_{k''_m m}$$

$$= \frac{(N_1 - p_1)! \cdots (N_n - p_n)!}{(N_1 - (m_1 + p_1))! \cdots (N_n - (m_n + p_n))!}$$

Other developments on the functions $f_{m_1,...,m_n}$ and $\hat{f}_{m_1,...,m_n}$ are identical to those of the multi-component classical case[8]. For instance, for the OMPP process, the distribution $\hat{f}_{m_1,...,m_n}(X''_{k''_1 1},\cdots,X''_{k''_m m})$ is usually replaced by the conditional m-tuplet distribution function $\hat{f}_{m_1,...,m_n}(X''_{k''_1 1},...,X''_{k''_m m}|X''_{i0})$ where the particle at location $X''_{i0}$ is said to be located at the origin and is of species i. Also, we have that:

$$\hat{f}_{m_1,...,m_{i-1},m_i+1,m_{i+1},...,m_n}(X''_{i0},X''_{k''_1 1},...,X''_{k''_m m}) = \hat{f}_1(X''_{i0})\hat{f}_{m_1,...,m_n}(X''_{k''_1 1},...,X''_{k''_m m}|X''_{i0})$$

where for homogeneous systems $\hat{f}_1(X''_{i0})$ is easily argued to yield[8] $\rho_i$. In particular, following the same arguments as for the classical case[8], partial m-tuplet distribution functions are related to PQNNPDF's for the GMPP process say as:

$$\hat{f}_{m_1,...,m_n}(X_{k_1 1},...,X_{k_m m}) = \sum_{s_1}\cdots\sum_{s_m} g^{QG}_{s_1,...,s_m}(X_{k_1 1},...,X_{k_m m}) \tag{34}$$

where $g^{QG}_{s_1,...,s_m}(X_{k_1 1},\cdots,X_{k_m m})$ is the marginal pdf for the event that the $s_1$-th nearest neighbor is of species $k_1$ at location $X_{k_1 1}$; … ; the $s_m$-th nearest neighbor is of species $k_m$ at location $X_{k_m m}$. The coordinates $X_{k_1 1},\cdots,X_{k_m m}$ are radially ordered, implying their radial portions are ordered as: $r_{k_1 1} < r_{k_2 2} < \cdots < r_{k_m m}$. Hence the coordinates $X_{k_1 1},\cdots,X_{k_m m}$ are generalized ordered. In the multiple sum of Eq. (34), $s_m$ varies from m to some large value, while $s_{m-1}$ varies from (m − 1) to ($s_m$ − 1),……, $s_1$ varies from 1 to ($s_2$ − 1). Usually, m does not have to be too large, implying $s_m$ does not have to vary to too large a value, thus further implying that the sum of Eq. (34) can be expected to usually be manageable enough to be performed accurately numerically (with the computer). This is understandable considering that large m values implies very distant neighbors, where the behavior of m-tuplet distribution functions are generally known to be "featureless". Hence large m values (and thus large $s_m$ values) are usually not considered. Also, the PQNNPDF in Eq. 34 is expected to get rapidly small beyond some values of $s_1$, … , $s_m$. It is therefore usual to assume m, $s_m$ << $N_i$. If $s_m$ were $\gtrsim N_i$, the species types that may be presumed for neighbors (ie, the $s_1$-th, … , $s_m$-th nearest neighbors, and other neighbors in-between the said nearest neighbors including the origin), would generally become difficult to reconcile with the values of $s_1$, … , $s_m$ in the sums of Eq. (34). $g^{QG}_{s_1,...,s_q}(X_{k_1 1},\cdots,X_{k_q q})$ may be constructed by appropriately integrating $g^{QG(l)}_{1,...,m}(X_{k_1 1},\cdots,X_{k_m m})$ for a select set of (l) and then adding up. (Clearly, we must have $s_1 < s_2 < \cdots < s_q \leq m$). The specific choices of the $k_1,\cdots,k_m$ set (as signified by (l) ) must be such that $k_{s_1} = k_1;\cdots;k_{s_q} = k_q$. Since there are ($s_1$ - 1) neighbors nearer to the origin than the $s_1$-th nearest neighbor, there are $n^{s_1-1}$ choices of species that can be made for the nearer neighbors. Similarly, there are $(s_2 - s_1 - 1)$ neighbors between the $s_1$-th and $s_2$-th nearest neighbors, and thus there are $n^{s_2-s_1-1}$ choices of species that can be made for such "in-between" choices. A similar thing can also be said for the $(s_3 - s_2 - 1)$ neighbors between the $s_2$-th and $s_3$-th nearest neighbors, etc. Hence the select set of $(l)$ indices indicated above will total (in number) $n^{\{(s_1-1)+(s_2-s_1-1)+...+(s_q-s_{q-1}-1)+(m-s_q)\}} = n^{m-q}$. It should be noted that because the coordinates are now generalized ordered, the distribution $\hat{f}_{m_1,...,m_n}(X_{k_1 1},...,X_{k_m m})$ will normalize to

$$\frac{N_1! \cdots N_n!}{(N_1 - m_1)! \cdots (N_n - m_n)!(m_1 + \cdots + m_n)!}.$$ The use of Eq. (34) to compute partial m-body distribution functions is expected to yield accurate results, thus providing a new and useful method amongst several others known in the literature[4,6] (which range from semi-classical methods to path integral methods and many more) for determining partial m-body distribution functions. Observe also that because PQNNPDF's can be determined for arbitrary system scale (as has earlier been discussed), partial m-body distribution functions can also be determined for arbitrary system scale employing Eq. (34).

# IV. Application To Weakly Correlated Quantum Systems

The scheme developed in sections II and III will now be applied mainly to quantum mixed systems under weak degeneracy where correlation functions deviate slightly from those of the classical "poisson" gas mixtures, while other thermodynamic regimes will be briefly discussed. Only results valid for macro-scale systems (thermodynamic limit) from sections II and III shall be employed. So far, we have implicitly assumed "short range" inter-particle interactions are in effect in the systems considered, of which many equilibrium multi-component systems belong. Such multi-component systems usually include those for which each particle species, when considered separately, actually involve "long range" inter-particle interactions; but when considered under equilibrium in a mixture with other particle species, we find for instance that in the presence of "overall charge neutrality", the effective inter-particle interactions is short ranged.[1]

The thermodynamic phase space is divided into four broad regions for the purpose of this paper, and each of the regions shall be considered in turn. The first region, which is the region of main application emphasis in this paper, involves weak degeneracy, stipulated as $\frac{1}{\rho} >> \lambda_i^3$. Also in this region, the density is assumed very low, making the average distance between particles much larger than the worst case effective short range ($r_0$) of inter-particle interactions. This implies that inter-particle interactions are largely weak, and this condition is broadly expressed as $\frac{1}{\rho} >> r_0^3$. In the weakly degenerate quantum mixed system, a reasonable zeroth order approximation for the Slater Sum is given as $W_{m_1,...,m_n}(X_{k_1 1},..., X_{k_m m}) \sim \exp(-\beta U_{m_1,...,m_n})$ (see section III), and thus expressions for PQNNPDF's may be well approximated by expressions for the PNNPDF's provided for classical multi-component systems[1]. For the ideal quantum fluid mixture that is weakly degenerate, the corresponding PNNPDF's become those of the classical poisson gas mixture[1]. The poisson result still remains a reasonable approximation for the weakly degenerate non-ideal quantum mixed system with sufficiently low density, where largely weak inter-particle interactions is in effect. Hence in this case (involving the first thermodynamic region of main interest), the normalization constant $h_{k_1 1,...,k_m m}$ in Eqs. (20) or (23) for the GMPP PQNNPDF for instance may be accurately approximated by the quantity $(\rho_1^{m_1} \cdots \rho_n^{m_n})$ which is applicable for the poisson fluid mixture.

Also, we have that in the weak interaction limit (applicable to the first thermodynamic regime currently under consideration), surface and shape effects are expected to be weak, and PQNNPDF's as formulated in section III may be expected to yield accurate results for all m (which include small m values such as m = 1 involving the first nearest neighbor). With the weak degeneracy weak interaction approximation for $g_{1,...,m}^{QG}(X_{k_1 1}, \cdots, X_{k_m m})$ for instance, the marginal pdf $g_{s_1,...,s_m}^{QG}(X_{k_1 1}, \cdots, X_{k_m m})$ can be constructed, and Eq. (34) can be used to provide the approximation for the partial m-tuplet distribution function $\hat{f}_{m_1,...,m_n}(X_{k_1 1},..., X_{k_m m})$ (applicable for the first thermodynamic regime) which should show slight deviation from that of the classical poisson gas mixture.

To determine the free energy, we begin by substituting the weak interaction, weak degeneracy approximation of Eq. 23 for $g^Q_{1,\ldots,m}(X_{k_1 1},\cdots,X_{k_m m})$ into Eq. (11) to get

$$P_j(X'_{j1}, v_1) = \sum_{(l)} \int_{(half\ space)} \cdots \int (\rho_{k_1}\cdots\rho_{k_m})\exp\left(-\frac{2}{3}\pi r_{k_m m}^3 \frac{p}{kT}\right) W_{m_1,\ldots,m_{j-1},(m_j+1),m_{j+1},\ldots,m_n}(X'_{j1}, X_{k_1 1},\ldots, X_{k_m m})\prod_{i=1}^{m} dX_{k_i i}$$

$$(m \gg 1)$$

(35)

In the weak degeneracy weak interaction limit, the product property of the multi-component Slater Sum as given in Eq. (4), applies for small m values, hence Eq. (35) can be rewritten employing m = 1 as:

$$P_j(X'_{j1}, v_1) = \sum_{k_1=1}^{n} \int_{(half-space)} \rho_{k_1} \exp\left(-\frac{2}{3}\pi r_{k_1 1}^3 \frac{p}{kT}\right) W_{m_{k_1},(m_j+1)}(X'_{j1}, X_{k_1 1})dX_{k_1 1} \quad (36)$$

Note that in the notation for $W_{m_{k_1},(m_j+1)}(X'_{j1}, X_{k_1 1})$, when $k_1 \neq j$, we have $m_j = 0$ and $m_{k_1} = 1$; and when $k_1 = j$, we have $m_j + 1 = 0$ and $m_{k_1} = 2$. In the absence of external forces, $W_{m_{k_1},(m_j+1)}(X'_{j1}, X_{k_1 1})$ depends only on the distance between the original particle situated at $X'_{j1}$, and its nearest neighbor[3], and thus we may write:

$$P_j(X'_{j1}, v_1) \approx \sum_{k_1=1}^{n} \int_0^{\pi}\int_0^{\pi}\int_0^{\infty} \rho_{k_1} r_{k_1 1}^2 \exp\left(-\frac{2}{3}\pi r_{k_1 1}^3 \rho\right) W_{m_{k_1},(m_j+1)}(r_{k_1 1})\sin\theta_{k_1 1} dr_{k_1 1} d\theta_{k_1 1} d\phi_{k_1 1}$$

$$= \sum_{k_1=1}^{n} 2\pi\rho_{k_1} \int_0^{\infty} r_{k_1 1}^2 \exp\left(-\frac{2}{3}\pi r_{k_1 1}^3 \rho\right) W_{m_{k_1},(m_j+1)}(r_{k_1 1})dr_{k_1 1}$$

(37)

Integrating this equation by parts yields:

$$P_j(X'_{j1}, v_1) \approx \sum_{k_1=1}^{n}\left[-\alpha_{k_1}\exp\left(-\frac{2}{3}\pi r_{k_1 1}^3\rho\right) W_{m_{k_1},(m_j+1)}(r_{k_1 1})\Bigg|_{r_{k_1 1}=0}^{\infty} + \alpha_{k_1}\int_0^{\infty}\exp\left(-\frac{2}{3}\pi r_{k_1 1}^3\rho\right) W'_{m_{k_1},(m_j+1)}(r_{k_1 1})dr_{k_1 1}\right]$$

(38)

where $W'_{m_{k_1},(m_j+1)}(r_{k_1 1}) = \dfrac{dW_{m_{k_1},(m_j+1)}(r_{k_1 1})}{dr_{k_1 1}}$. In the weak degeneracy limit,

$W_{m_1,\ldots,m_n}(X_{k_1 1},\ldots, X_{k_m m})$ behaves approximately as $\exp(-\beta U_{m_1,\ldots,m_n})$, and because of "large" inter-particle distances providing for largely weak interactions, $\beta U_{m_1,\ldots,m_n}$ is "almost always" close to zero, and so $W_{m_1,\ldots,m_n}(X_{k_1 1},\ldots, X_{k_m m})$ is close to unity often enough. From Eq. (11) and the normalization condition for PQNNPDF's, this means $P_j(X'_{j1}, v_1)$ is also close to unity. With the system behaving classically, and under weak interactions, ε is close to unity and may be written as $\varepsilon = 1 +$ (terms of order $\rho_1, \ldots, \rho_n$, and higher); hence we readily see that we may write the following approximation from Eq. (15).

$$\varepsilon \approx \sum_{i=1}^{n} \alpha_i P_i \qquad (39)$$

We note that:

$$\frac{\partial \alpha_j}{\partial \rho_i} = \begin{cases} -\dfrac{\alpha_j}{\rho} & (i \neq j) \\ \dfrac{(1-\alpha_i)}{\rho} & (i = j) \end{cases} \qquad (40)$$

Hence, using Eq. (39), we have that:

$$\rho_i \frac{\partial \varepsilon}{\partial \rho_i} = \rho_i \sum_{j(j \neq i)} \alpha_j \left( \frac{\partial P_j}{\partial \rho_i} - \frac{P_j}{\rho} \right) + \frac{P_i \rho_i}{\rho}(1-\alpha_i) + \rho_i \alpha_i \frac{\partial P_i}{\partial \rho_i} = \rho_i \sum_{j} \alpha_j \frac{\partial P_j}{\partial \rho_i} - \alpha_i (\varepsilon - P_i)$$

But we had that $\varepsilon \approx P_i \approx 1$; hence we have that

$$\rho_i \frac{\partial \varepsilon}{\partial \rho_i} \approx \rho_i \sum_{j} \alpha_j \frac{\partial P_j}{\partial \rho_i} \qquad (41)$$

From Eqs. (38) and (40), we have:

$$\frac{\partial P_j(X'_{j1}, v_1)}{\partial \rho_i} = \sum_{k_1(k_1 \neq i)} \left( -\frac{\alpha_{k_1}}{\rho} W_{m_{k_1},(m_j+1)}(0) \right) + \frac{(1-\alpha_i)}{\rho} W_{m_i,(m_j+1)}(0)$$

$$+ \sum_{k_1} \alpha_{k_1} \int_0^\infty \left( -\frac{2}{3} \pi r_{k_1 1}^3 \right) \exp\left( -\frac{2}{3} \pi r_{k_1 1}^3 \rho \right) W'_{m_{k_1},(m_j+1)}(r_{k_1 1}) dr_{k_1 1}$$

$$+ \sum_{k_1(k_1 \neq i)} \left( -\frac{\alpha_{k_1}}{\rho} \right) \int_0^\infty \exp\left( -\frac{2}{3} \pi r_{k_1 1}^3 \rho \right) W'_{m_{k_1},(m_j+1)}(r_{k_1 1}) dr_{k_1 1}$$

$$+ \frac{(1-\alpha_i)}{\rho} \int_0^\infty \exp\left( -\frac{2}{3} \pi r_{i1}^3 \rho \right) W'_{m_i,(m_j+1)}(r_{i1}) dr_{i1}$$

$$= \sum_{k_1} \alpha_{k_1} \int_0^\infty \left( -\frac{2}{3} \pi r_{k_1 1}^3 \right) \exp\left( -\frac{2}{3} \pi r_{k_1 1}^3 \rho \right) W'_{m_{k_1},(m_j+1)}(r_{k_1 1}) dr_{k_1 1}$$

(42)

We may assume the integral $\int_0^\infty \left( \frac{2}{3} \pi r_{k_1 1}^3 \right) W'_{m_{k_1},(m_j+1)}(r_{k_1 1}) dr_{k_1 1}$ not only exists, but also has a finite range ($0 \leq r_{k_1 1} \leq R$ say) over which it has its major contribution. In which case, if the density is so low that we may approximate the exponential terms in Eq. (42) as $\approx 1$ for $0 \leq r_{k_1 1} \leq R$, we then write in the low density limit

$$\frac{\partial P_j(X'_{j1}, v_1)}{\partial \rho_i} = -\sum_{k_1} \alpha_{k_1} \int_0^\infty \left(\frac{2}{3}\pi r_{k_1 1}^3\right) W'_{m_{k_1},(m_j+1)}(r_{k_1 1}) dr_{k_1 1} \qquad (43)$$

Employing Eq. (43) in Eq. (41), and assuming $\varepsilon \approx 1$, we get in the weak correlation, weak degeneracy, low density limit of a quantum mixed system, the following equation of state (using Eq. 21):

$$\frac{\phi}{\rho} = 1 + \rho \sum_j \sum_l \alpha_j \alpha_l \int_0^\infty \left(\frac{2}{3}\pi r_{l1}^3\right) W'_{m_l,(m_j+1)}(r_{l1}) dr_{l1} \qquad (44)$$

The multiple sum of Eq. (44) is the second virial coefficient of quantum mixed systems. The second virial coefficient may also be written as[4]:

$$-\frac{1}{2}\sum_j \sum_l \alpha_j \alpha_l \int \left(W_{m_l,(m_j+1)}(r_{l1}) - 1\right) d^3 r_{l1} = -\sum_j \sum_l \alpha_j \alpha_l \int_0^\infty 2\pi r_{l1}^2 \left(W_{m_l,(m_j+1)}(r_{l1}) - 1\right) dr_{l1}$$

(45)

By integrating by parts, Eq. (45) is readily shown to coincide with the multiple sum of Eq. (44) employing the result that $\lim_{r_{l1} \to \infty} W_{m_l,(m_j+1)}(r_{l1}) \to 1$.

We now briefly discuss the remaining broad thermodynamic regions. The second thermodynamic regime involves largely weak inter-particle interactions and medium/strong degeneracy ($\frac{1}{\rho} \lesssim \lambda_i^3; \frac{1}{\rho} \gg r_0^3$). This is the regime of much recent activities whereby neutral Fermionic and Bosonic atoms (including their mixtures) are trapped and cooled to temperatures in the nanokelvin range. Such dilute, weakly interacting, highly degenerate systems[7] have been found to be readily amenable to thorough investigations, and this has therefore allowed several quantum and condensed matter theories to be brought under intense scrutiny. In our current analysis, we do not consider external/lattice forces which are usually introduced by the confining magnetic/optical fields for the system. The present scheme of this paper however may be extended largely along the same line as was outlined for classical systems[1,2] to include external forces thereby allowing for a more detailed investigation. Unlike the first thermodynamic regime considered, the mild form of the product property of the Slater Sum may generally not be satisfied for small m values in the second thermodynamic regime. However, the weak inter-particle interactions implies that as a first approximation, we may employ the ideal quantum gas mixture expression for $W_{m_1,\ldots,m_n}$. The Hamiltonian operator is separable for each component of the mixture, hence we may write for instance $W^0_{N_1,\ldots,N_n} = W^0_{N_1} W^0_{N_2} \cdots W^0_{N_n}$ (the superscript "0" indicating an ideal quantum gas mixture). Following the method given in the book by D. ter Haar[4] (see section 4 of chapter 8 of the text), we may therefore rewrite Eq. 2 (with generalized ordered coordinates) as:

$$W^0{}_{N_1,\ldots,N_n} = W^0{}_{N_1} W^0{}_{N_2} \cdots W^0{}_{N_n}$$

$$= \prod_{i=1}^{n} \left[ \sum_{p_i} \varepsilon_{p_i} \exp\left\{ -\frac{m_i}{2\hbar^2 \beta} \left( \left|X_{i1} - X_{i(p_i1)}\right|^2 + \left|X_{i2} - X_{i(p_i2)}\right|^2 + \cdots + \left|X_{iN_i} - X_{i(p_iN_i)}\right|^2 \right) \right\} \right]$$

(46)

$\sum_{p_i}$ is a summation over all $N_i!$ permutations of particle coordinates of the i-th species; $(p_ik)$ is a permutation of the k; $\varepsilon_{p_i}$ is +1 for Bose-Einstein statistics for the i-th species presumed to be Bosonic, while for Fermi-Dirac statistics, it is equal to +1 or -1 according to whether the permutation of particle coordinates is even or odd.

The first term of the permutation sum ($\sum_{p_i}$) is the identity permutation where no coordinate is permuted, thus yielding the value unity. This term is clearly the largest of all terms in the permutation sum, and Eq. 46 may be rewritten as:

$$W^0{}_{N_1,\ldots,N_n} = \prod_{i=1}^{n} \left[ 1 \pm \left( \sum_{\substack{permute \\ 1-pair}} \exp(\cdots)_i \right)_{N_i} + \left( \sum_{\substack{permute \\ 2-pairs}} \exp(\cdots)_i \right)_{N_i} \pm \left( \sum_{\substack{permute \\ 3-pairs}} \exp(\cdots)_i \right)_{N_i} + \cdots \right]$$

$$= \prod_{i=1}^{n} \left( \sum_{k} \varepsilon_k \left( \sum_{\substack{permute \\ k-pairs}} \exp(\cdots)_i \right)_{N_i} \right)$$

(47)

The subscript $N_i$ after the equality signs indicates the number of particles subjected to the permutation process as indicated within the corresponding brackets. The + sign applies for Bosons and the − sign applies for Fermions, while

$$\exp(\cdots)_i = \exp\left\{ -\frac{m_i}{2\hbar^2 \beta} \left( \left|X_{i1} - X_{i(p_i1)}\right|^2 + \cdots + \left|X_{iN_i} - X_{i(p_iN_i)}\right|^2 \right) \right\}$$

$\varepsilon_k$ = +1 for Bosons, and for Fermions, $\varepsilon_k$ = +1 for k even (ie. even number of pairs permuted), and $\varepsilon_k$ = -1 for k odd (ie. odd number of pairs permuted). The zeroth order approximation for PQNNPDF's (GMPP case) in the second thermodynamic regime under consideration becomes:

$$g^{QG0}_{1,\ldots,m}\left(X_{k_1 1},\ldots,X_{k_m m}\right) = h_{k_1 1,\ldots,k_m m}\left[\prod_{i=1}^{n}\left(\sum_{k}\varepsilon_k\left(\sum_{\substack{permute\\k-pairs}}\exp(\cdots)_i\right)_{m_i}\right)\right]\exp\left(-\frac{4}{3}\pi r^3_{k_m m}\frac{p}{kT}\right)$$

$$(m \gg 1)$$

(48)

The superscript "0" in $g^{QG0}_{1,\ldots,m}$ implies ideal quantum gas.

Because of the need of the use of the mild form of the product property of the Slater Sum in developing PQNNPDF's (see section III), we find that the condition m >> 1 is required for Eq. 48 to be valid in the medium/strong degeneracy region even for weak inter-particle interactions. Hence Eq. 48 is not readily used in a purely analytic scheme (like was done to compute the second virial coefficient in the first thermodynamic regime) to obtain the statistical parameter $P_i$ or $<P_i>$ (see section II). Hence we must embark on an elaborate computer numerical scheme (which we do not currently address) to compute $P_i$ or $<P_i>$ and iteratively solve Eq. 15 (as indicated in section II) to obtain accurate results for ε or the free energy of the system.

The third thermodynamic regime involves medium/strong inter-particle interactions and weak degeneracy ($\frac{1}{\rho} \lesssim r_0^3; \frac{1}{\rho} \gg \lambda_i^3$) applicable for most liquid systems for instance. We already remarked that weak quantum effects in this regime implies that the Slater Sum can be approximated as the Boltzmann's factor. Hence PQNNPDF's may be approximated as those governing classical multi-component systems.[1] A requirement of the condition m >> 1 is rather more apparent in this case as it quickly becomes a poor approximation to assume that the particle at the origin may interact only with its first nearest neighbor. Each particle in the system must now be presumed to interact effectively with several nearest neighbor particles (probably at most a few tens or so), and developments in section III show that by choosing m large enough (m >> 1) the product property of the Slater Sum can be used to lead to valid expressions for PQNNPDF's. A similar remark is also applicable for the fourth thermodynamic regime involving medium/strong inter-particle interactions, and medium/strong degeneracy ($\frac{1}{\rho} \lesssim r_0^3; \frac{1}{\rho} \lesssim \lambda_i^3$) applicable for instance for solid state systems in general. Hence like the second thermodynamic regime, investigation of the third and fourth thermodynamic regimes require elaborate computer numerical schemes (which we do not address at this time) if investigation of the properties of various condensed state of matter must be done accurately.

# v. Conclusions and Remarks

The success of the concept of NNPDF's (first extended to PNNPDF's and QNNPDF's, and now to PQNNPDF's, along with the notions of generalized order and the one particle phase space) in the investigation of general classical systems[1,2] as well as single component quantum systems,[3] have once again been demonstrated, this time for the most general type of material systems, quantum mixtures. Application has been restricted mainly to the thermodynamic limit of systems that are weakly degenerate, with largely weak inter-particle interactions, where the second virial coefficient has been successfully reproduced for multi-component quantum systems. It is expected that similar success can be achieved using the same formalism for moderate to strongly correlated / interacting many-body quantum systems and other thermodynamic regimes at other length scales.

As has been noted in Ref, 3, the present formalism has general applicability and validity for arbitrary material systems including those for which particle number cannot be fixed (as in grand-canonical ensemble formalism) or where "observable quantities" are more conveniently addressed employing representations other than the coordinate representation. Also, $W_{m_1,...,m_n}(X_{k_1 1},...,X_{k_m m})$ (and related functions) dealt with in this paper, bear close analogy with other functions encountered in field theoretic methods of physics (see section I). The approach of this paper essentially reduces a many-body problem to a "few-body" problem, whereby the most computationally intensive aspect of the scheme employs numerical calculations for a subsystem involving a few nearest neighbor particles to determine the statistical parameter <$P_j$>. The scheme then provides an "extrapolation" of the results for <$P_j$> for the few-body subsystem to results for systems of arbitrary scale. In which case, results for $W_{m_1,...,m_n}(X_{k_1 1},...,X_{k_m m})$, and <$P_j$> (which are results for a finite particle cluster), provide a "signature" from which properties at other system scales may be inferred. Issues relating to the range of inter-particle interaction, or the effect of external forces in the evaluation of the free energy and structure of quantum systems may be investigated largely along the same line as was outlined for classical systems[1,2].

As remarked in Ref. 3, extreme difficulties exist in accurately determining quantities such as $W_{m_1,...,m_n}(X_{k_1 1},...,X_{k_m m})$ (for all m), but it is in the case of small m values (the case which is of importance in the method of this paper), that we may hope substantial progress may be made especially by means of numerical computation. A major point of this paper however, is that, the method developed greatly reduces the degree of complexity of the quantum statistical thermodynamic problem.

The strong analogy between the scheme of this paper and that used for classical multi-component systems was possible because of the product property of $W_{m_1,...,m_n}(X_{k_1 1},...,X_{k_m m})$ which plays the role of the Boltzmann's factor for classical systems. It is our intention to continue to explore the product property which has so far received little attention in the literature especially for quantum mixed systems.

# Acknowledgments

This work was supported by the University of Wisconsin-Madison.